\documentclass[12pt]{article}
\usepackage{amsfonts,amsmath,amssymb}
\begin{document}
\begin{center} {\bf VISCOUS BRANE COSMOLOGY WITH A  BRANE-BULK
ENERGY INTERCHANGE TERM} \vspace{1cm}

 Iver
Brevik\footnote{Department of Energy and Process Engineering,
Norwegian University of Science and Technology, N-7491 Trondheim,
Norway; e-mail: iver.h.brevik@ntnu.no}, Jon-Mattis
B{\o}rven\footnote{Department of Energy and Process Engineering,
Norwegian University of Science and Technology, N-7491 Trondheim,
Norway}, and Sebastian Ng\footnote{Department of Energy and
Process Engineering, Norwegian University of Science and
Technology, N-7491 Trondheim, Norway}
\end{center}

\bigskip

\begin{abstract}
We assume a flat brane located at $y=0$, surrounded by an AdS
space, and consider the 5D Einstein equations when the energy flux
component of the energy-momentum tensor is related to the Hubble
parameter through a constant $Q$. We calculate the metric tensor,
as well as the Hubble parameter on the brane, when $Q$ is small.
As a special case, if the brane is tensionless, the influence from
$Q$ on the Hubble parameter is absent. We also consider the
emission of gravitons from the brane, by means of the Boltzmann
equation. Comparing the
 energy conservation equation derived herefrom
with the energy conservation equation for a viscous fluid on the
brane, we find that the entropy change for the fluid in the
emission process has to be negative. This peculiar effect is
related to the fluid on the brane being a non-closed thermodynamic
system. The negative entropy property for non-closed systems is
encountered in other areas in physics also, in particular, in
connection with the Casimir effect at finite temperature.

\end{abstract}

\bigskip
KEY WORDS: Brane cosmology; viscous cosmology; Randall-Sundrum;
gravitons.

\section{Introduction}

When considering brane world perturbative cosmology one is
confronted with a plethora of phenomena, among which some are
unknown even with respect to sign. For instance, as discussed
recently by Durrer \cite{durrer05}, during ordinary inflation
gravitational waves are generated. For a given inflationary
potential their amplitudes can be calculated. In a brane world
context, a fraction of these waves will be radiated from the brane
into the bulk and thereby reduce the gravitational wave amplitude.
On the other hand there may also be gravitational waves generated
in the bulk, and some of these may accumulate on the brane,
increasing the amplitude of gravitational waves on the brane.
Thus, depending on the circumstances, even the sign of the brane
world effect on a gravitational wave background is unknown.

In view of this state of affairs it may seem desirable to allow
for realistic, non-ideal, properties of the cosmic fluid on the
brane. Therewith one may hope to get some guidance in restricting
the number of possibilities in the description of physical
processes. One natural option in this direction is
 to allow for a {\it bulk viscosity}. As
is known from ordinary hydromechanics, one is easily led astray in
many cases if one ignores the viscosity effects. A bulk viscosity,
in contrast to a shear viscosity, is compatible with the
assumption about complete isotropy of the cosmic fluid. We shall
assume in the following that there is a constant bulk viscosity
$\zeta$ present, but shall ignore the shear viscosity. We shall
work in terms of Gaussian normal coordinates and assume that there
is one single brane present at fixed position $y=0$. That is, we
adopt essentially the Randall-Sundrum type II model
\cite{randall99}, although this model is strictly speaking
non-cosmological. (It might in this context seem natural, as an
alternative, to introduce a spherical model in which the brane is
an expanding surface mimicking the cosmological expansion. This
implies a 5D cosmological solution of Einstein's equations with a
negative cosmological constant. Theories of this kind have been
worked out in Ref.~\cite{kraus99} and \cite{ida00}; cf. also the
recent review in Ref.~\cite{langlois05}.)

Dissipative cosmology theories were worked out some years ago -
cf., for instance, the reviews  \cite{gron90,maartens95} - whereas
the theory of viscous fluids in a brane context was recently
investigated in Refs.~\cite{chen01,harko03,brevik04}. The presence
of a dissipative fluid on the brane gives us the possibility to
compare with general thermodynamical principles; in particular,
the behaviour of entropy in irreversible processes.

An important point in the present context is that we will relax
the condition about zero energy flux from the brane, $T_{ty}=0$,
in the $y$ direction. This means physically that we draw into
consideration the production of gravitons.
 Since the emission of gravitons into the bulk can be described
via the Boltzmann equation, this is the case that we will be
henceforth  interested in (we thus do not consider any further the
absorption of gravitons on the brane).
 We shall describe this interchange effect in
a simple way phenomenologically, by introducing a non-vanishing
energy flux component $T_{ty}$. It is only this component of the
5D energy-momentum tensor $T_{AB}$ that comes into play in the
present context.

In section 3 we will be concerned with the energy conservation
equation for the viscous fluid on the brane. Comparing with the
corresponding equation derived from the Boltzmann equation, it
actually turns out that the emission process corresponds to a
negative entropy change for the thermodynamic subsystem on the
brane. This may be  an unexpected result, but it does not simply
run into conflict with basic thermodynamics all the time that the
thermodynamic principles apply only to a closed system; in our
case this means  the brane fluid plus the bulk particles.

\section{Einstein's Equations, with an Interchange Term}

As mentioned, we assume that there is one single brane located at
$y=0$. We take the spatial curvature $k$ to be zero.  The metric
will be taken in the form
\begin{equation}
ds^2= -n^2(t,y)dt^2+a^2(t,y)\delta_{ij}dx^idx^j +dy^2. \label{1}
\end{equation}
The quantities $n(t,y)$ and $a(t,y)$ are determined from
Einstein's equations  which are, with $\Lambda$ the 5D
cosmological constant,
\begin{equation}
R_{AB}-\frac{1}{2}g_{AB}R +g_{AB}\Lambda=\kappa^2T_{AB}. \label{2}
\end{equation}
Here the coordinate indices are numbered as
$x^A=(t,x^1,x^2,x^3,y)$, with $\kappa^2=8\pi G_5$  the 5D
gravitational coupling. With the metric (\ref{1}) Einstein's
equations in a coordinate basis have been worked out before
\cite{binetruy00,binetruy00a,brevik02,brevik04,brevik04a}, but it
is convenient to give them also here for reference purposes. When
$k=0$,
\begin{equation}
 3\left\{ \left( \frac{\dot{a}}{a}\right)^2 -n^2
\left[\frac{a''}{a} +\left(\frac{a'}{a} \right)^2 \right]
\right\}-\Lambda n^2=\kappa^2 T_{tt}, \label{3}
\end{equation}
\[
a^2\delta_{ij} \Bigg\{ \frac{a'}{a}\left(
\frac{a'}{a}+\frac{2n'}{n}\right) +\frac{2a''}{a}+\frac{n''}{n} \]
\begin{equation}
 +\frac{1}{n^2}\Big[
\frac{\dot{a}}{a}\left(-\frac{\dot{a}}{a}+\frac{2\dot{n}}{n}\right)-\frac{2\ddot{a}}{a}\Big]
+ \Lambda \Bigg\} =\kappa^2 T_{ij}, \label{4}
\end{equation}
\begin{equation}
 3\left(
\frac{\dot{a}}{a}\frac{n'}{n}-\frac{\dot{a}'}{a}
\right)=\kappa^2T_{ty}, \label{5}
\end{equation}
\begin{equation}
3\Bigg\{\frac{a'}{a}\left(\frac{a'}{a}+\frac{n'}{n}\right)-\frac{1}{n^2}\Big[\frac{\dot{a}}{a}\left(
\frac{\dot{a}}{a}-\frac{\dot{n}}{n}\right)
+\frac{\ddot{a}}{a}\Big]\Bigg\}+\Lambda=\kappa^2 T_{yy}. \label{6}
\end{equation}
Here overdots and primes mean derivatives with respect to $t$ and
$y$, respectively. The energy-momentum tensor is taken in the form
\begin{equation}
T_{AB}=\delta(y)\left[-\sigma g_{\mu\nu}+\rho U_\mu
U_\nu+\tilde{p}\,h_{\mu\nu}\right]\delta_A^\mu \delta_B^\nu,
\label{7}
\end{equation}
where $h_{\mu\nu}=g_{\mu\nu}+U_\mu U_\nu$ is the projection tensor
and $\tilde{p}=p-3H_0\zeta$ is the effective pressure,
$H_0=\dot{a}_0/a_0$ being the Hubble parameter on the brane $y=0$.
As gauge condition we take ${n_0}(t)=1$, which physically means
that the proper time on the brane is equal to the cosmological
time coordinate. As $n_0(t)$ is a constant, the scalar expansion
$\theta= {U^\mu}_{;\mu}=3H_0+\dot{n}_0/n_0$ reduces to $3H_0$
\cite{brevik04}. The energy-momentum expression (\ref{7}) is
composed of two parts: one part which in an orthonormal frame
means $T_{tt}=\delta(y) \sigma,\, T_{ij}=-\delta(y) \sigma
\delta_{ij}$, which is in accordance with the equation of state
$p=-\rho$ for a cosmic brane \cite{vilenkin81}, and there is a
second part describing the energy-momentum for a viscous fluid. We
work henceforth in an orthonormal frame, where $U^\mu=(1,0,0,0)$,
and let generally the subscript zero be referring to the brane.

Consider next the junction conditions applied to Eqs.~(\ref{3})
and (\ref{4}) at $y=0$. For the distributional parts we have
\cite{brevik04}
\begin{equation}
\frac{[a']}{a_0}=-\frac{1}{3}\kappa^2 (\sigma+\rho), \label{8}
\end{equation}
\begin{equation}
[n']=\frac{1}{3}\kappa^2(-\sigma+2\rho+3\tilde{p}), \label{9}
\end{equation}
where $[a']=a'(0^+)-a'(0^-)$, and similarly for $[n']$.

For the nondistributional parts we have
\begin{equation}
\left(
\frac{\dot{a}}{na}\right)^2-\frac{a''}{a}-\left(\frac{a'}{a}\right)^2=\frac{1}{3}\Lambda,
\label{10}
\end{equation}
\[ \frac{a'}{a}\left( \frac{a'}{a}+\frac{2n'}{n}\right)+\frac{2a''}{a}+\frac{n''}{n} \]
\begin{equation}
+\frac{1}{n^2}\Big[ \frac{\dot{a}}{a} \left(
-\frac{\dot{a}}{a}+\frac{2\dot{n}}{n} \right)
-\frac{2\ddot{a}}{a}\Big] = -\Lambda. \label{11}
\end{equation}
We now turn attention to the energy flux component $T_{ty}$,
attempting to model it in a simple way. It seems natural to assume
that the energy flux from  the brane was stronger in the early
stages of the universe when the Hubble parameter $H=\dot{a}/a$ was
large,  than it is today. As ansatz we shall adopt a simple
proportionality, i.e., $T_{ty}=-QH$, where $Q$ is a constant. This
ansatz  actually leads to mathematical simplifications also.
Namely, in (\ref{5}) we can eliminate $\dot{a}/a$ to get the
equation
\begin{equation}
\frac{n'}{n}-\frac{\dot{a}'}{\dot{a}}=-\frac{1}{3}\kappa^2Q
\label{12}
\end{equation}
which, after integration with respect to $y$, yields
\begin{equation}
n(t,y)=\frac{\dot{a}(t,y)}{\dot{a}_0(t)}\,{\rm
e}^{-\kappa^2Q|y|/3}, \label{13}
\end{equation}
where the $Z_2$ symmetry $y\rightarrow -y$ is taken into account.
Note that the condition $n_0=1$ is obeyed, and that a positive
value of $Q$ leads to a decreasing value of the metric component
$n(t,y)$ with increasing distances $|y|$ from the brane.

Consider next equation (\ref{10}). After some algebra and use of
(\ref{13}) we can write it in the form
\begin{equation}
\frac{d}{dy}\left\{
\left(\frac{\dot{a}a}{n}\right)^2-(aa')^2-\frac{1}{6}\Lambda a^4
\right\}=\frac{2}{3}\kappa^2 Q a^2 \dot{a}_0^2\,{\rm
e}^{2\kappa^2Q|y|/3}. \label{14}
\end{equation}
A nonvanishing value of $Q$ thus spoils the conservation of the
expression between the curly parentheses and thereby changes the
conventional 5D brane version of Friedmann's first equation.

We shall not solve (\ref{14}) in general, but limit ourselves to
the case when $Q$ is small. Specifically, we shall consider the
condition (assumed here that $Q>0$)
\begin{equation}
\kappa^2 Q |y| \ll 1. \label{15}
\end{equation}
Stated in another way: we consider only distances $|y|$ from the
brane for which (\ref{15}) is satisfied. This region close to the
brane is obviously also the one of main physical interest. Our
calculation below will go only to the first order in $Q$.

When the exponential in (\ref{14}) is replaced with unity, we need
only the expression for $a(t,y)$ on the right hand side  that
pertains to the case $Q=0$. For $y=0$ we then have the equation
\begin{equation}
\left(\frac{\dot{a}_0}{a_0}\right)^2=\frac{1}{6}\Lambda+\frac{\kappa^4}{36}(\sigma
+\rho)^2+\frac{C}{a_0^4}, \label{16}
\end{equation}
where $C=C(t)$ is an integration constant with respect to $y$. As
$\rho=\rho(t)$, this equation can be solved for $a_0$ only in
special cases. Let us give the explicit solution when $\rho=0$
\cite{brevik04a}:
\begin{equation}
a_0(t; \rho=0)= \frac{1}{2\sqrt{\lambda}f(t)}\left[f^4(t)-4\lambda
C\right]^{1/2}, \label{17}
\end{equation}
where the constant
\begin{equation}
\lambda=\frac{1}{6}\Lambda+\frac{1}{36}\kappa^4\sigma^2 \label{18}
\end{equation}
can be interpreted as an effective four-dimensional cosmological
constant in the five-dimensional theory, and
\begin{equation}
f(t)={\rm e}^{\sqrt{\lambda}(t+c_0)}, \label{19}
\end{equation}
$c_0$ being a new integration constant (recall that $k=0$ is
assumed).

Now considering the $Q=0$ solution for $a(t,y)$ away from the
brane, we shall assume only the AdS case, i.e., $\Lambda <0$. We
then have, for general $y$,
\[ a^2(t,y;Q=0)= \frac{1}{2}a_0^2\Big[
1+\frac{\kappa^4}{6\Lambda}(\sigma+\rho)^2\Big]+\frac{3C}{\Lambda
\,a_0^2} \]
\[ +\left\{ \frac{1}{2}a_0^2\Big[1-\frac{\kappa^4}{6\Lambda}(\sigma+\rho)^2\Big]-\frac{3C}{\Lambda\,a_0^2}\right\}
\cosh(2\mu\,y) \]
\begin{equation}
-\frac{\kappa^2}{6\mu}(\sigma+\rho)a_0^2\sinh(2\mu\,|y|),
\label{20}
\end{equation}
where $\mu=\sqrt{-\Lambda/6}$. (Note that the condition (\ref{15})
does not necessarily imply that the argument $(2\mu y)$ is small.)
The terms containing the quantity $C$ are not of main interest
here and will hereafter be omitted (the extra term $C/a_0^4$ in
the Friedmann equation (\ref{16}) is called the "radiation term").

The expression (\ref{20}) is easily integrated with respect to
$y$, and so the whole equation (\ref{14}) can be integrated. Again
omitting a "radiation" type term we obtain
\[ \left(\frac{\dot{a}}{na}\right)^2=\frac{1}{6}\Lambda
+\left(\frac{a'}{a}\right)^2+\frac{2}{3}\kappa^2
Q\,\frac{\dot{a}_0^2}{a^4}\,\Bigg\{ \frac{1}{2}a_0^2
\left(1+\frac{\kappa^4 \sigma^2}{6\Lambda }\right)y \]
\begin{equation}
+\frac{1}{4\mu}a_0^2\left( 1-\frac{\kappa^4
\sigma^2}{6\Lambda}\right)\sinh 2\mu y -\frac{\kappa^2
\sigma}{12\mu^2}\, a_0^2\cosh 2\mu y \Bigg\}. \label{21}
\end{equation}
Note that it is equivalent here whether we insert $a_0^2(t;Q=0)$
or the more general expression $a_0^2 (t;Q) \equiv a_0^2(t)$ on
the right hand side, all the time that we work to the first order
in $Q$.

The expression (\ref{21}) can be evaluated on the brane. Let us
take $y=0^+$; then $a_0'/a_0=-\frac{1}{6}\kappa^2(\rho+\sigma)$.
Recalling that $n_0(t)=1$ we get
\begin{equation}
H_0^2=\frac{\lambda+\frac{1}{18}\kappa^4\sigma
\rho+\frac{1}{36}\kappa^4\rho^2} {1+\frac{\kappa^4}{18\mu^2}\sigma
Q}. \label{22}
\end{equation}
This is our main result. It shows how Friedmann's first equation
becomes modified in the presence of a nonvanishing $Q$. When
$Q=0$, (\ref{22}) reduces to (28) in Ref.~\cite{brevik04} (when
the radiation term is omitted).

We note the following points:

1)  There is no influence from the viscosity in (\ref{22}). This
arises from the fact that Friedmann's first equation refers to the
energy in the fluid, not to the pressure, and it is only in the
latter context that viscosity plays a role (cf. $p\rightarrow
\tilde{p}$ above).

2)  If $\sigma>0$ (i.e., a positive tensile stress on the brane),
and if $Q>0$, then the magnitude of the Hubble parameter becomes
{\it diminished} by the presence of $Q$.

3)  If the brane is tensionless, $\sigma=0$, there is no influence
from $Q$ on the Hubble parameter at all, irrespective of the value
of of $\rho$.

What is the {\it sign} of the constant $Q$? Although we have not
discussed this issue in detail,  it is fairly obvious from the
expression (\ref{13}) that in order for the present model to be
physically reasonable one should have  $Q>0$.  The influence from
$T_{ty}$ on the brane has to decay with increasing distances
 $|y|$ from the brane.

 In the next section we shall consider the energy conservation equation for
 the fluid on the brane. This equation can be derived from the
 Boltzmann equation describing the emission process, or alternatively from the
 standard conservation equation for a viscous fluid.
 Comparison of the equations  will enable us to discuss the behaviour of entropy.

\section{Radiating Brane. Energy Conservation Equation in the
Presence of Viscosity}

A reasonable physical model for the brane-bulk interaction is to
assume that bulk gravitons are produced by fluctuations of brane
matter. Assuming as before an AdS bulk, we can take the gravitons
to be created by the collision of pairs of particles on the brane.
The process can be described in various ways. Let us briefly
review here the kind of approach advocated by Langlois et al.
\cite{langlois02,langlois03,langlois04}. The process can be
described as
\begin{equation}
\psi \bar{\psi} \rightarrow g, \label{23}
\end{equation}
where $\psi$ is a standard model particle and $g$ is a graviton.
The equation of state is written in the conventional form
\begin{equation}
p=w \rho, \label{24}
\end{equation}
where $w=1/3$ for highly relativistic particles. One can make use
of the Boltzmann equation on the brane,
\begin{equation}
\dot{\rho}+3H_0(\rho+p)=-\int \frac{d^3p}{(2\pi)^3}\,C[f],
\label{25}
\end{equation}
where the collision term is
\begin{equation}
C[f]=\frac{1}{2}\int
\frac{d^3p_1}{(2\pi)^32E_1}\frac{d^3p_2}{(2\pi)^3 2E_2}\,\sum
|M|^2f_1f_2(2\pi)^4\delta^{(4)}(p_1+p_2-p). \label{26}
\end{equation}
Here $f$ is the distribution function for gravitons, and $M$ is
the scattering amplitude. The calculation leads to the following
result, with $w=1/3$,
\begin{equation}
\dot{\rho}+4H_0\,\rho=-\frac{315}{512\pi^3}\hat{g}\,\kappa^2\,T^8,
\label{27}
\end{equation}
when the matter is in thermal equilibrium at a temperature $T$.
Here $\hat{g}=(2/3)g_s+g_f+4g_v=188.7$, where $g_s=4,\, g_f=90,\,
g_v=24$ refer to the degrees of freedom for scalars, fermions, and
vectors, respectively, in the standard model.

We now go back to the formalism of the previous section.
Evaluating (\ref{12}) on both sides of the brane, and inserting
the expressions (\ref{8}) and (\ref{9}) for the jumps $[a']$ and
$[n']$, we obtain the energy conservation equation in the
following form:
\begin{equation}
\dot{\rho}+3H_0(\rho+p)=9\zeta H_0^2. \label{28}
\end{equation}
In this equation $Q$ does not appear explicitly, but its influence
is hidden in the $Q-$dependent expression for $H_0$; cf.
(\ref{22}). It is notable that (\ref{28}) has precisely the same
form as in ordinary viscous four-dimensional cosmology
\cite{gron90,brevik94,brevik02a}, although there is seemingly no
simple physical reason why this should be so.

Let us compare equations (\ref{27}) and (\ref{28}). When $w=1/3$
their left hand sides are the same, but on the right hand sides
there is a striking difference in that the {\it signs} are
opposite. As we know from  from ordinary thermodynamics the bulk
viscosity $\zeta$ (as well as the shear viscosity $\eta$) are
taken  to be positive quantities; this arising from the condition
that the entropy change in an irreversible process for a closed
system is positive \cite{landau87}.

So, the following conclusion naturally emerges: The emission of
gravitons into the bulk, as described by (\ref{27}), is
accompanied by a {\it negative} entropy change in the cosmic fluid
on the brane. It corresponds to a negative $\zeta$. The negativity
of the entropy change is counterintuitive, but does not violate
thermodynamics as one would be inclined to conclude at first. The
reason is that the fluid on the brane forms a {\it non-closed}
thermodynamic system; in order to close the system one has to
include the particles in the bulk also. General relationships,
such as the law about entropy increase in an irreversible process,
applies to a closed thermodynamic system only.

This particular  effect is not so uncommon after all. It becomes
natural here to compare with the theory of the Casimir effect.
Imagine that there are two parallel metal plates, separated by a
fixed gap $a$ of the order of $1\,\mu$m, at a temperature $T$. At
low $T$, there exists a finite temperature interval in which the
Casimir free energy $F$ is {\it increasing}  with increasing
values of $T$, keeping $a$ constant. This corresponds to a
negative entropy $S=-\partial F/\partial T$ in the actual
temperature interval. The reason is evidently that these Casimir
quantities are not concerned with the thermodynamic quantities of
the  closed system, but only with the interaction part of it. And
the Casimir force is derived just from the interaction part of the
free energy. We have considered this effect repeatedly earlier
\cite{hoye03,brevik04b,brevik05,hoye05}, and the theory has been
corroborated by other works considering the Casimir effect from
different viewpoints - cf. for instance Sernelius et al.
\cite{sernelius04,bostrom04,sernelius05} - although it is only
fair to say that a full consensus on this point has not so far
been achieved in the literature.

\section{Summary}

Our starting point was the metric (\ref{1}), corresponding to zero
spatial curvature, whereby the 5D Einstein equations (\ref{2})
took the form (\ref{3})-(\ref{6}). On the $y=0$ brane, endowed
with a constant tension $\sigma$, a fluid with density $\rho$ and
constant bulk viscosity $\zeta$ was assumed, corresponding to the
energy-momentum tensor in the form of (\ref{7}).

The main results of the present paper are:

$\bullet$  Assuming the energy flux component to satisfy the
proportionality $T_{ty}=-QH$ with $Q$ a constant, we found the
metric component $n(t,y)$ to be given by (\ref{13}). This
expression shows that $Q>0$ in order for the present kind of
theory to be meaningful; the  influence from the  brane-bulk
interaction is expected to decay for increasing values of $|y|$.

$\bullet$ In the limit of small $Q$, the metric component $a(t,y)$
is determined from (\ref{21}), which in turn leads to the
expression (\ref{22}) for the Hubble parameter $H_0$ on the brane.
This expression shows, in particular, that the influence from $Q$
on $H_0$ is absent if the brane is tensionless, $\sigma=0$.

$\bullet$ The emission of gravitons can be described through the
Boltzmann equation on the brane
\cite{langlois02,langlois03,langlois04}. The corresponding energy
conservation equation (\ref{27}), when compared with the energy
conservation equation (\ref{28}) for a viscous fluid on the brane,
shows that $\zeta$ has to be negative, corresponding to a negative
entropy change. This counterintuitive effect relates to the fact
that the fluid on the brane is a non-closed thermodynamic system.
The negative entropy effect has parallels in other areas of
physics also, notably in connection with the Casimir effect at a
finite temperature
\cite{hoye03,brevik04b,brevik05,hoye05,sernelius04,bostrom04,sernelius05}.



\begin{thebibliography}{99}
\bibitem{durrer05}
Durrer, R. (2005). hep-th/0507006.
\bibitem{randall99}
Randall, L. and Sundrum, R. (1999). {\it Phys. Rev. Lett.} {\bf
83}, 4690.
\bibitem{kraus99}
Kraus, P. (1999). {\it JHEP} {\bf 9912}, 011.
\bibitem{ida00}
Ida, D. (2000). {\it JHEP} {\bf 0009}, 019.
\bibitem{langlois05}
Langlois, D. (2005) hep-th/0509231.
\bibitem{gron90}
Gr{\o}n, {\O}. (1990). {\it Astrophys. Space Sci.} {\bf 173}, 191.
\bibitem{maartens95}
Maartens, R. (1995). {\it Class. Quant. Grav.} {\bf 12}, 1455.
\bibitem{chen01}
Chen, C.-M., Harko, T. and Mak, M. K. (2001). {\it Phys. Rev. D}
{\bf 64}, 124017.
\bibitem{harko03}
Harko, T. and Mak, M. K. (2003). {\it Class. Quant. Grav.} {\bf
20}, 407.
\bibitem{brevik04}
Brevik, I. and Hallanger, A. (2004). {\it Phys. Rev. D} {\bf 69},
024009.
\bibitem{binetruy00}
Bin$\acute{\rm e}$truy, P.,  Deffayet, C., Ellwanger, U.,and
Langlois, D. (2000). {\it Phys. Lett. B} {\bf 477}, 285.
\bibitem{binetruy00a}
Bin$\acute{\rm e}$truy, P.,  Deffayet, C., and  Langlois, D.
(2000). {\it Nucl. Phys. B} {\bf 565}, 269.
\bibitem{brevik02}
 Brevik, I., Ghoroku, K., Odintsov, S. D., and Yahiro, M. (2002). {\it Phys.
Rev. D} {\bf 66}, 064016.
\bibitem{brevik04a}
Brevik, I., B{\o}rkje, K., and Morten, J. P. (2004). {\it Gen.
Rel. Grav.} {\bf 36}, 2021.
\bibitem{vilenkin81}
Vilenkin, A. (1981). {\it Phys. Rev. D } {\bf 23}, 852.
\bibitem{langlois02}
Langlois, D., Sorbo, L., and Rodriguez-Martinez, M. (2002). {\it
Phys. Rev. Lett.} {\bf 89}, 171301.
\bibitem{langlois03}
Langlois, D. and Sorbo, L. (2003). {\it Phys. Rev. D} {\bf 68},
084006.
\bibitem{langlois04}
Langlois, D. (2004). astro-ph/0403579.
\bibitem{brevik94}
Brevik, I. and Heen, L. T. (1994). {\it Astrophys. Space Sci.}
{\bf 219}, 99.
\bibitem{brevik02a}
Brevik, I. (2002). {\it Phys. Rev. D}  {\bf 65}, 127302.
\bibitem{landau87}
Landau, L. D. and Lifshitz, E. M. (1987). {\it Fluid Mechanics,
2nd ed.}. Pergamon Press, Oxford, Section 49.
\bibitem{hoye03}
H{\o}ye, J. S., Brevik, I., Aarseth, J. B. and Milton, K. A.
(2003). {\it Phys. Rev. E} {\bf 67}, 056116.
\bibitem{brevik04b}
Brevik, I., Aarseth, J. B., H{\o}ye, J. S. and Milton, K. A.
(2004). {\it Proc. 6th Workshop on Quantum Field Theory Under the
Influence of External Conditions}, ed. K. A. Milton. Rinton Press,
New Jersey, p.~54 [ quant-ph/0311094].
\bibitem{brevik05}
Brevik, I., Aarseth, J. B., H{\o}ye, J. S. and Milton, K. A.
(2005). {\it Phys. Rev. E} {\bf 71}, 056101.
\bibitem{hoye05}
H{\o}ye, J. S.,  Brevik, I., Aarseth, J. B. and Milton, K. A.
 (2005).  quant-ph/0506025 v2.
\bibitem{sernelius04}
Sernelius, Bo E. and Bostr{\"o}m, M.  (2004). {\it Proc. 6th
Workshop on Quantum Field Theory Under the Influence of External
Conditions}, ed. K. A. Milton. Rinton Press, New Jersey, p.~82.
\bibitem{bostrom04}
Bostr{\"o}m, M. and Sernelius, Bo E. (2004). {\it Physica A} {\bf
339}, 53.
\bibitem{sernelius05}
Sernelius, Bo E. (2005). {\it Phys. Rev. B} {\bf 71}, 235114.







\end{thebibliography}
\end{document}